\documentclass[aps,showpacs,prd,twocolumn]{revtex4}
\usepackage{eurosym}
\usepackage{amssymb}
\usepackage{amsfonts}
\usepackage{amsmath}
\usepackage{graphicx}
\usepackage{color}
\usepackage[dvips]{epsfig}
\usepackage[dvips]{graphicx}
\usepackage{float}

\newcommand{\be}{\begin{equation}}
\newcommand{\ee}{\end{equation}}
\input{tcilatex}
\begin{document}

\title{Trapping of Spin-0 fields on tube-like topological defects}
\author{R. Casana$^{1}$, A. R. Gomes$^{2,3}$, R. Menezes$^{4,5}$, F. C. Simas%
$^{1}$}
\affiliation{$^1$ Departamento de F\'\i sica, Universidade Federal do Maranh\~ao,
65080-805, S\~ao Lu\'\i s, Maranh\~ao, Brasil\\
$^3$ Departamento de F\' isica, Instituto Federal de Educa\c c\~ao,
Ci\^encia e Tecnologia do Maranh\~ao (IFMA), 65025-001, S\~ao Lu\'\i s,
Maranh\~ao, Brasil\\
$^4$ Departamento de Ci\^encias Exatas, Universidade Federal da Para\'\i ba,
58297-000, Rio Tinto, Para\'\i ba, Brasil\\
$^5$ Departamento de F\' isica, Universidade Federal de Campina Grande,
58109-970, Campina Grande, Para\'\i ba, Brasil}

\begin{abstract}
We have considered the localization of resonant bosonic
states  described by a scalar field $\Phi$ trapped in tube-like topological
defects. The tubes are formed by radial symmetric defects in $(2,1)$
dimensions, constructed with two scalar fields $\phi$ and $\chi$, and
embedded in the $(3,1)-$dimensional Minkowski spacetime.  The general
coupling between the topological defect and the scalar field $\Phi$ is given
by the potential $\eta F(\phi,\chi)\Phi^2$. After a convenient decomposition
of the field $\Phi$, we find that the amplitudes of the radial modes satisfy
Schr\"odinger-like equations whose eigenvalues are the masses of the bosonic
resonances. Specifically, we have analyzed two simple couplings: the first
one is $F(\phi,\chi)=\chi^2$ for a fourth-order potential and, the second
one is a sixth-order interaction characterized by $F(\phi,\chi)=(\phi\chi)^2$%
. In both cases the Schr\"odinger-like equations are numerically solved with
appropriated boundary conditions. Several resonance peaks for both models
are obtained and the numerical analysis showed that the fourth-order
potential generates more resonances than the sixth-order one.
\end{abstract}

\maketitle


\section{introduction}

Topological defects are full of interesting realizations in physics. As
examples one can cite applications in the study of quark confinement \cite%
{phenom}, gravitation and cosmology \cite{cosm} and condensed matter physics
\cite{condmat}. The idea that we live in a multidimensional brane world \cite%
{brane-origins} has been applied to the issue of topological defects, with
interesting insights to the problem of cosmological constant and hierarchy
\cite{br1,br2,br3,br4,br5}. The inclusion of scalar fields is connected to
the concept of dynamically generated thick branes \cite%
{thick1,thick2,thick3,thick4,thick5,thick6,thick7,thick8,thick9,thick10,thick_rev}
with interesting results concerning to their structure, collision properties
and localization of fields \cite{field-localiz}.

In this work we study a class of topological defects embedded in a flat
spacetime. Despite the absence of gravity, the defects are in a way similar
to $(5,1)$ branes, in the sense of being the result of an embedding of a
topological defect in two extra dimensions. Specifically we consider $(2,1)$%
-dimensional topological defects with two coupled real scalar fields,
constructed within a simplified model describing a color dielectric medium.%
\textbf{\ }We follow and extend the procedure presented in Refs. \cite%
{bmm1,bmm2} for the case of two real scalar fields. We show that the
embedding of the radial defect in $(3,1)$ -dimensions form a tube-like
topological defect and study its structure. In the present work, we show
that the construction with two scalar fields is crucial for studying
processes of field localization of spin-0 particles with particular interest
for resonance effects.

There are several examples of two-dimensional topological defects with
radial symmetry. For instance, vortices are $(2,1)$ dimensional soliton
solutions in $U(1)$ gauge theory with complex scalars and the Higgs
mechanism \cite{vortex}. Recently, interesting extensions for non-Abelian
fields have been constructed \cite{nonabelian}. Ring-type objects are a
subject of great interest (see, for instance, the review \cite{rv}). An
important example of solutions of this type in curved spacetime are black
rings, which are solutions of general relativity with extra dimensions (see,
for instance, \cite{er}). Another\ interesting example of ring solitons in
gauge field theory are the anomalous solitons, where the gauge field is
constrained after fixing its Chern-Simons number by including fermions into
the system \cite{rt,rub}. Also in non-Abelian gauge field theory we can have
smooth, finite energy loops stabilized by the magnetic energy, forming
non-abelian rings. Considering the non-Abelian Yang-Mills-Higgs theory for
the group $SU(2)$, ring solitons were obtained in \cite{kks} which are more
general solutions than the magnetic monopoles \cite{tp}. Other constructions
give sphaleron rings \cite{kkl}. A sphaleron is a static unstable solution
of the classical equations of motion; it is a saddle point configuration
separating topologically distinct vacua \cite{rub2}. Despite their possible
instability, the sphaleron rings can perhaps have an importance as mediators
of baryon number violating processes \cite{km,rv}.

In this work we are interested in the Abelian version of the color
dielectric model \cite{fried,wil} with two real scalar fields without
fermions. The topological defect formed is neither a ring nor a vortex (in
the sense that there is no winding number). A first physical motivation is
to show that a dielectric model with two coupled real scalar fields can be
responsible for an effective breaking of translational invariance, leading
to a topological defect capable of trapping scalar particles with simple
couplings. In this way, this work can be inserted in a series of previous
studies of trapping of spin-0 fields for specific topological defects. In
the context of braneworlds there are several examples of such analysis. In $%
(5,1)$ dimensions (one extra dimension) see, for instance, the works in \cite%
{scalar-localiz}.  The construction of brane models with one extra
dimension can be done starting from the solution in $(1,1)$ dimensions with
the help of the first-order formalism \cite{thick1}. The first-order
formalism is very helpful for achieving explicit analytical solutions for
brane models. As far as we know, brane models with two extra dimensions have
no similar formalism developed yet, which turns explicit solutions much more
difficult to obtain. In this way another motivation for this work can be
inserted in the searching for such first-order formalism. Indeed, here we
consider a $(3,1)-$dimensional system where the two extra dimensions are
described by the usual $(y,z)$ coordinates, but where the application of a
first-order formalism was done only in a flat Minkowski spacetime,
neglecting gravity effects. 

The manuscript is presented in the following way: In Section II we consider $%
(2,1)$-dimensional radial topological defects embedded in a $(3,1)$%
-dimensional flat spacetime. In Section III, we study some aspects of
localization of scalar fields in this system, and numerically investigate
resonance effects in Section IV. Our conclusions are presented in Section V.

\section{A TUBE IN (3, 1)-DIMENSIONS}

\label{tube} We start with an Abelian version of the color dielectric model
\cite{fried,wil} without fermions. We can express this by the following
action
\begin{eqnarray}
S &=&\int dtdxdydz\biggl[\frac{1}{2}\partial _{M}\phi \partial ^{M}\phi +%
\frac{1}{2}\partial _{M}\chi \partial ^{M}\chi \\
&&-\frac{g(\phi ,\chi )}{4}F_{MN}F^{MN}-eA_{M}J^{M}\biggr],
\label{fundamental}
\end{eqnarray}%
where $g(\phi ,\chi )$ is the electric permittivity. We use capital letters $%
M$, $N$ for all $(3,1)$ dimensions (coordinates $(t,x,y,z)$) and Greek
letters $\mu ,\nu $ for $(1,1)$ dimensions (coordinates $(t,x)$). We
particularize to $\displaystyle J_{M}=\frac{1}{r}(\delta (r),0,0,0)$, with $%
r=\sqrt{y^{2}+z^{2}}$, representing the charge density. The factor $1/r$
comes from expressing the delta function $\delta (\vec{r})$ in cylindrical
coordinates.

The equations of motion for static and radial scalar fields are
\begin{eqnarray}
\nabla ^{2}\phi &=&-\frac{1}{2}g_{\phi }E^{2},  \label{nablaphi} \\
\nabla ^{2}\chi &=&-\frac{1}{2}g_{\chi }E^{2},  \label{nablachi}
\end{eqnarray}%
where in this paper we use the simplified notation $f_{\phi }=\partial
f/\partial \phi $, $f_{\phi \chi }=\partial ^{2}f/(\partial \phi \partial
\chi )$ and similar constructions for other derivatives of a differentiable
function $f$. The Maxwell equations, $\partial _{M}(gF^{MN})=eJ^{N}$ lead to
the following expression for the electric field
\begin{equation}
E=\frac{e}{rg}\,\,.
\end{equation}%
We consider a medium with electric permittivity given by (similar relation
for one field was found in \cite{bmm1})
\begin{equation}
g(\phi ,\chi )=\frac{e^{2}}{2V(\phi ,\chi )},  \label{gV}
\end{equation}%
where
\begin{equation}
V(\phi ,\chi )=\frac{1}{2}(W_{\phi }^{2}+W_{\chi }^{2})
\end{equation}%
is a potential that generates a two-field topological defect in $(1,1)$
dimensions and $W(\phi ,\chi )$ is the superpotential. This means that we
are considering a dielectric model where the electric permittivity is
infinity at the vacua of the potential $V(\phi ,\chi )$. In $(1,1)$
dimensions the vacua of the potential are located at $\pm \infty $. However,
we are constructing a model in $(3,1)$ dimensions with radial symmetry. We
will see then that one of the vacua occurs in $r=0$, meaning a divergent
dielectric permittivity along the center of the tube. We can also say that
the charge density given by $J^{M}$ polarizes the vacuum, leading to a
nontrivial topology.

With Eqs. (\ref{gV}), (\ref{nablaphi}) and (\ref{nablachi}), the equations
of motion for the scalar fields are
\begin{eqnarray}
\frac{1}{r}\frac{d}{dr}\biggr(r\frac{d\phi }{dr}\biggr) &=&-\frac{e^{2}}{%
2r^{2}}\frac{g_{\phi }}{g^{2}}  \label{eqmotionphi} \\
&=&\frac{1}{r^{2}}(W_{\phi }W_{\phi \phi }+W_{\chi }W_{\chi \phi })  \notag
\end{eqnarray}%
\begin{eqnarray}
\frac{1}{r}\frac{d}{dr}\biggr(r\frac{d\chi }{dr}\biggr) &=&-\frac{e^{2}}{%
2r^{2}}\frac{g_{\chi }}{g^{2}}  \label{eqmotionchi} \\
&=&\frac{1}{r^{2}}(W_{\phi }W_{\phi \chi }+W_{\chi }W_{\chi \chi }).  \notag
\end{eqnarray}

Following a similar procedure established in \cite{bmm1} for one-field
defects, for a given superpotential $W$ it can be shown that the solutions
of the first-order equations,
\begin{equation}
\frac{d\phi }{dr}=\frac{1}{r}W_{\phi }~\ ,~\ \frac{d\chi }{dr}=\frac{1}{r}%
W_{\chi },  \label{eqfirstorderchi}
\end{equation}%
are also solutions of the second-order equations (\ref{eqmotionphi}) and (%
\ref{eqmotionchi}).

Therefore, the tube in $(3,1)$-dimension is effectively
described by the action 
\begin{equation}
S_{tube}=\!\int \!\!dtd^{3}x\left( \frac{1}{2}\partial _{M}\phi \partial
^{M}\phi +\frac{1}{2}\partial _{M}\chi \partial ^{M}\chi -U(\phi ,\chi
)\right) ,  \label{actionring}
\end{equation}%
with
\begin{equation}
U(\phi ,\chi )=\frac{1}{2r^{2}}(W_{\phi }^{2}+W_{\chi }^{2}),
\end{equation}%
leading to the same equations of motion for the fields $\phi (r)$ and $\chi
(r)$ given by Eqs. (\ref{eqmotionphi}) and (\ref{eqmotionchi}).

The explicit dependence of $r=\sqrt{y^{2}+z^{2}}$ follows closely and
generalizes for two fields the construction of \cite{bmm1,bmm2} for evading
Derrick-Hobarts' theorem \cite{raj1,raj2,raj3}. The explicit breaking of
translational invariance in the action is present in some scenarios of QCD
\cite{color1a,color1b,color2a,color2b,deconf}, brane intersections \cite%
{br-inta,br-intb}, noncommutative field theory \cite{n-coma,n-comb} and
condensed matter physics \cite{dobr,bor}.

In this work we consider the following superpotential \cite{bsr}
\begin{equation}
W(\phi ,\chi )=\lambda \left( \phi -\frac{1}{3}\phi ^{3}-s\phi \chi
^{2}\right) ,  \label{superpotential}
\end{equation}%
corresponding to the potential
\begin{equation}
U(\phi ,\chi )=\frac{\lambda ^{2}}{r^{2}}\biggl(\frac{1}{2}(1-\phi
^{2}-s\chi ^{2})^{2}+2s^{2}\phi ^{2}\chi ^{2}\biggr),
\end{equation}%
which generates kink-like and lump-like solutions given, respectively, by
(more general solutions for this model where found in Ref. \cite{igg})
\begin{eqnarray}
\phi (r) &=&\frac{\left( \frac{r}{r_{0}}\right) ^{4\lambda s}-1}{\left(
\frac{r}{r_{0}}\right) ^{4\lambda s}+1}\,\mathrm{\ },  \notag \\[-0.2cm]
&&  \label{solring} \\[-0.2cm]
\chi (r) &=&\pm 2\sqrt{\frac{1}{s}-2}\,\frac{\left( \frac{r}{r_{0}}\right)
^{2\lambda s}}{\left( \frac{r}{r_{0}}\right) ^{4\lambda s}+1}.  \notag
\end{eqnarray}

\begin{figure}[]
\hspace{-0.15cm}\includegraphics[{angle=0,width=4.3cm}]{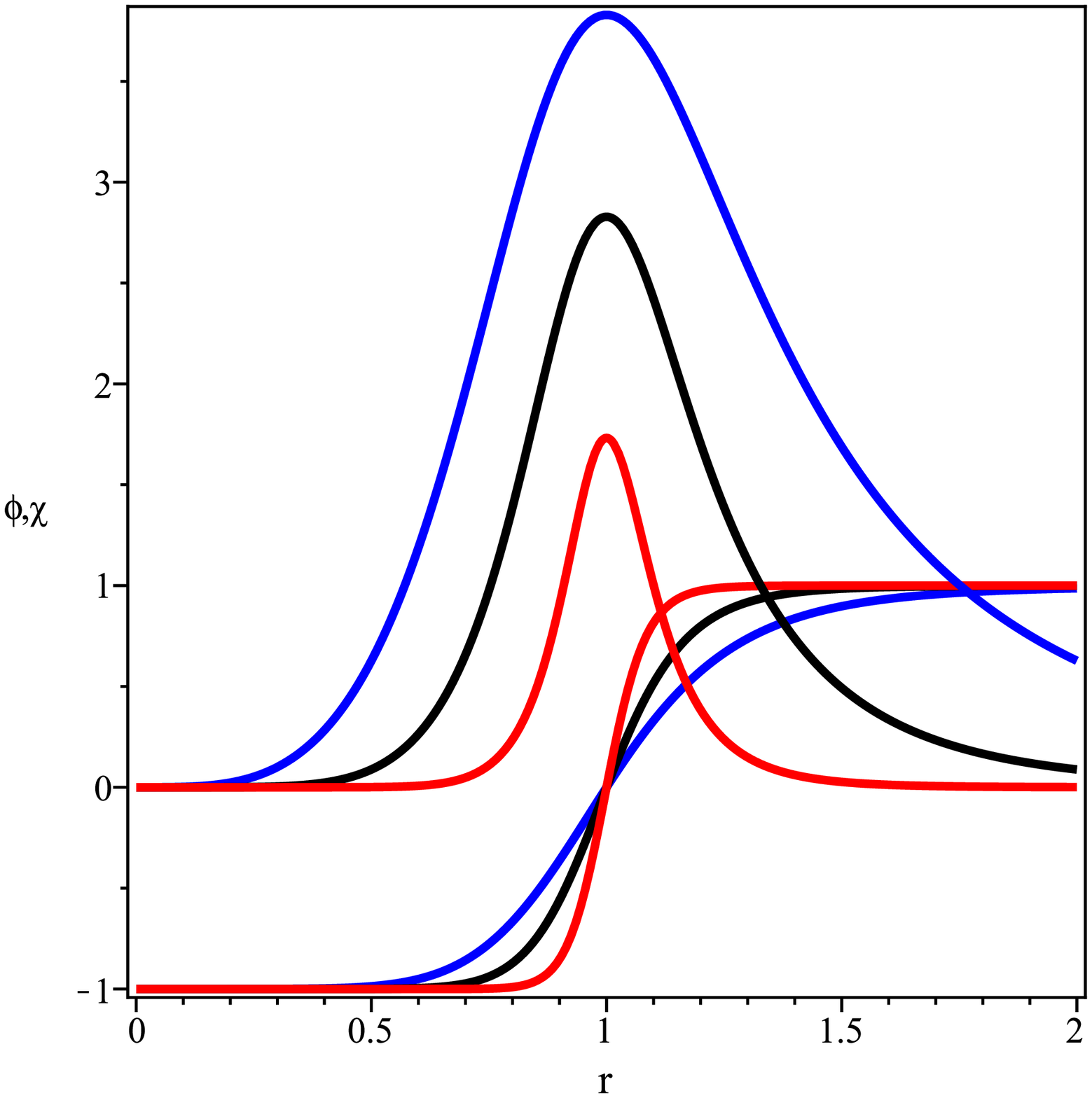}%
\includegraphics[{angle=0,width=4.3cm}]{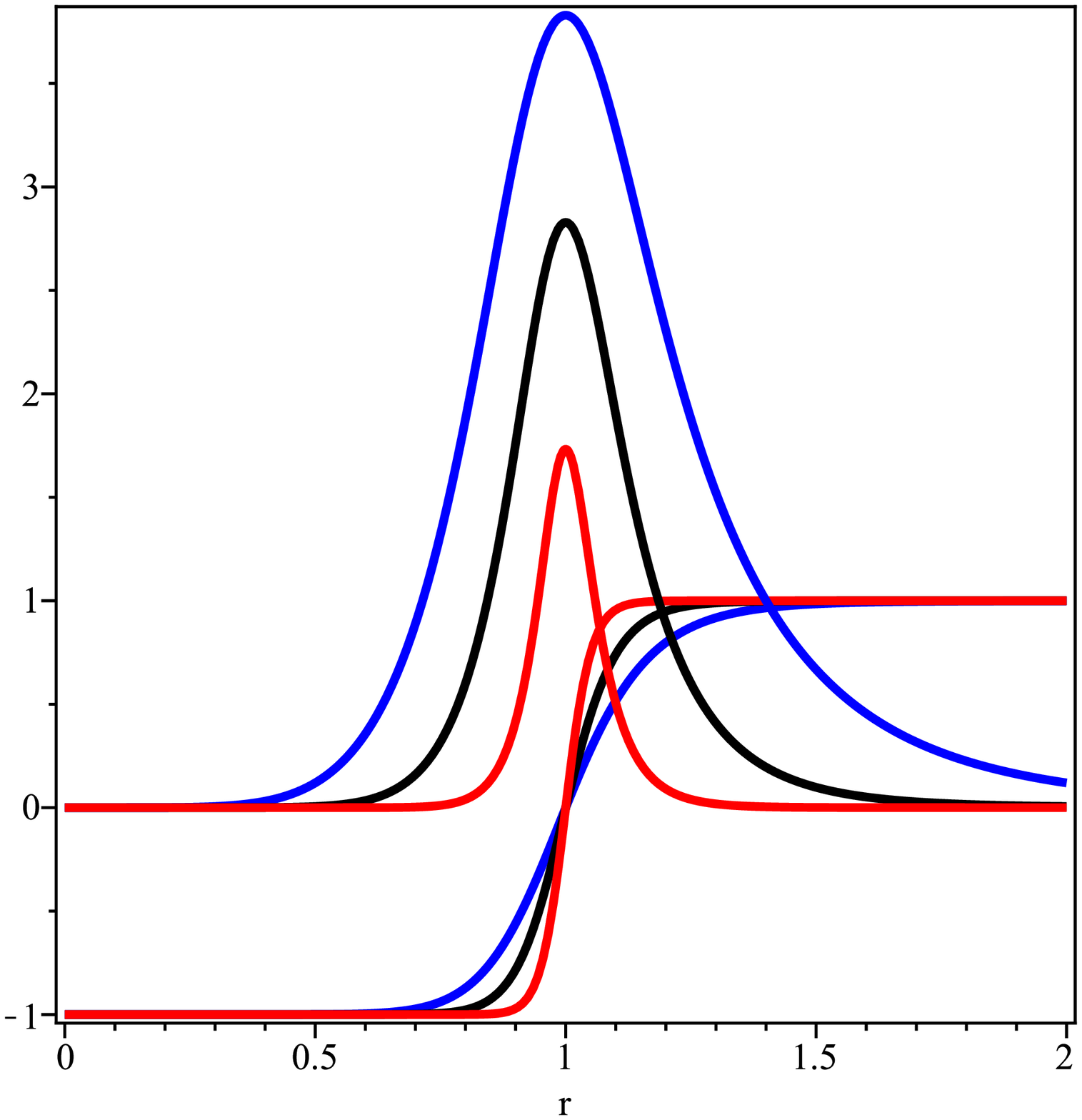}\vspace{-0.25cm}
\caption{The functions $\protect\phi(r)$ (kink-like) and $\protect\chi(r)$
(lump-like). We fix $r_0=1$. We have a) $\protect\lambda=30$ (left) and b) $%
\protect\lambda=50$ (right). In all figures $s=0.06$ (blue), $s=0.1$
(black), $s=0.2$ (red). }
\label{field}
\end{figure}

With the condition $0<s<1/2$, the solutions provide a radial profile with $%
r_{0}$ identified with the radius of the tube's cross section. The profiles
for the $\phi ,\chi $ fields are shown in Fig. \ref{field}. There it is
shown that for increasing $\lambda $ the defect becomes narrower. Also it is
verified that for small values of $s$, the effects of the field $\chi $ are
more predominant than those of $\phi $ and we have a wider defect. On the
other hand, larger values of $s$ show that the $\chi $ field is responsible
for the process of generating a thicker tube. In particular, the limit $%
s=1/2 $ recovers the one-field $\phi $ limit and a result from the
literature (the solution for $p=1$ in the notation of ref. \cite{bmm1}) is
recovered.

In $(1,1)$ dimensions the potential given by $(W_{\phi }^{2}+W_{\chi
}^{2})~/2 $ has minima at $(\pm 1,0)$ and $(0,\pm \sqrt{1/s})$ with $s>0$;
static solutions from the equations of motion connect the minima $(\pm 1,0)$
and form Bloch walls \cite{bsr}, which are defects with an internal
structure given by the $\chi$ field. The limit $s\rightarrow 0.5$ turns the
two-field problem into a one-field model with solution known as Ising wall.
In the present work the presence of the field $\chi $ also contributes to
generate an internal structure to the tube formed. Later in this work we
will see that this is crucial for localizing scalar fields with both fourth-
and sixth-order potentials.

The energy density of the $2$-dimensional radial defect is
\begin{eqnarray}
T_{00} &=&8\left( \frac{\lambda s}{r_{0}}\right) ^{2}\frac{\left( \frac{r}{%
r_{0}}\right) ^{4\lambda s-2}}{\left( \frac{r}{r_{0}}\right) ^{4\lambda s}+1}
\label{density} \\
&&~\times \left\{ 4\left( \frac{r}{r_{0}}\right) ^{4\lambda s}+\left( \frac{1%
}{s}-2\right) \left[ \left( \frac{r}{r_{0}}\right) ^{4\lambda s}-1\right]
^{2}\right\}  \notag
\end{eqnarray}%
Here we consider $T_{00}$ finite in $r=0$, which restricts the parameters to
satisfy $\lambda s\geq \frac{1}{2}$ when $\lambda >1$.

For fixed $\lambda >1$ and $\frac{1}{2\lambda }\leqslant s<\frac{1}{2}$, $%
T_{00}(r)$ changes from a lump centered in $r=0$ $\left( s=\frac{1}{2\lambda
}\right) $ to a peak centered around $r_{0}$\ $\left( s=\frac{1}{2}\right) $%
. For larger values of $\lambda $ and lower values of $s$, the contribution
of the $\chi$ field is higher and the defect appears as a thick tube
structure whose center is localized between the origin and $r_{0}$. On the
other hand, for larger values of $\lambda $ and larger values of $s$, the
defect looks like as a thin tube centered around $r_{0}$, and the field $%
\phi $ has the stronger contribution to the energy density.

The total energy in the $yz-$plane is given by $E=8\pi \lambda /3$, which
can be identified with the mass of the $(2,1)$ radial topological defect,
that is, the total mass (per unit length of the x-direction) of the tube.

\section{Spin-0 localization}

We consider a scalar field $\Phi $ in a region where exists a radial defect
constructed with the scalar fields $\phi ,\chi $. In the present analysis we
neglect the backreaction on the defect by considering that the interaction
between the scalar fields is sufficiently weak in comparison to the
self-interaction that generates the defect. In the following, we designate $%
\Phi $ as the weak field, and $\phi ,\chi $ the strong ones. We write the
action describing the system as
\begin{equation}
S_{1}=\int dtdxdydz\left( \frac{1}{2}\partial _{M}\Phi \partial ^{M}\Phi -%
\frac{\eta }{2}F\left( \phi ,\chi \right) \Phi ^{2}\right) ,  \label{Sscalar}
\end{equation}

The \ equation of motion of the scalar field is
\begin{equation*}
\partial _{\mu }\partial ^{\mu }\Phi -\nabla _{T}^{2}\Phi +\eta F\left( \phi
,\chi \right) \Phi =0
\end{equation*}%
where$~$%
\begin{equation*}
\partial _{\mu }\partial ^{\mu }=\square =\frac{\partial ^{2}}{\partial t^{2}%
}-\frac{\partial ^{2}}{\partial x^{2}}~\ ,~\ \ \nabla _{T}^{2}=\frac{%
\partial ^{2}}{\partial y^{2}}+\frac{\partial ^{2}}{\partial z^{2}}
\end{equation*}%
We consider a coupling $F\left( \phi ,\chi \right) =F\left( r\right) $ and
require $\mathcal{\eta }F(r)\rightarrow 0$ for $0\sim r\ll r_{0}$ and for $%
r\gg r_{0}$. It is possible to decompose the field $\Phi (t,x,y,z)$ as

\begin{equation}
\Phi (t,x,y,z)=\sum_{\ell ,n}\varphi _{\ell n}(t,x)\rho _{\ell n}\left(
r\right) e^{i\ell \theta },  \label{phirho}
\end{equation}%
where $\ell =0,1,2,...$ is related to the angular momentum eigenvalue. The
set $\left\{ e^{i\ell \theta }\right\} $ is orthogonal in $\theta \in
\lbrack 0,2\pi ]$. \ The field $\varphi _{\ell n}(t,x)$ satisfies the
two-dimensional Klein-Gordon equation%
\begin{equation}
\left( \square +m_{\ell n}^{2}\right) \varphi _{\ell n}=0
\end{equation}%
and the amplitude $\rho _{n\ell }(r)$ satisfies the radial Schr\"{o}%
dinger-like equation\
\begin{equation}
-\rho _{n\ell }^{\prime \prime }-\frac{1}{r}\rho _{n\ell }^{\prime }+V\left(
r\right) \rho _{n\ell }(r)=m_{\ell n}^{2}\rho _{n\ell }(r),  \label{sch2D}
\end{equation}%
for fixed $\ell $, where the Schr\"odinger potential is given by%
\begin{equation}  \label{pot_sch}
V=\frac{\ell ^{2}}{r^{2}}+\mathcal{\eta }F.
\end{equation}

By requiring that Eq. (\ref{sch2D}) defines a self-adjoint differential
operator in $r\in $\thinspace $\lbrack 0,+\infty )$, the Sturm-Liouville
theory \ establishes the orthonormality condition for the components $\rho
_{n\ell }(r)$,
\begin{equation}
\int_{0}^{\infty }drr\rho _{n\ell }(r)\rho _{m\ell }(r)=\delta _{mn}.
\label{orth}
\end{equation}

The action given by Eq. (\ref{Sscalar}) can be integrated in the $(y,z)$
dimensions, leading to
\begin{equation}
S_{1}=\int dtdx\sum_{\ell n}\left( \frac{1}{2}\partial _{\mu }\varphi _{\ell
n}\partial ^{\mu }\varphi _{\ell n}-\frac{1}{2}m_{\ell n}^{2}\varphi _{\ell
n}\right) .  \label{S_1}
\end{equation}%
This shows that the field $\varphi _{\ell n}$ is a massive two-dimensional
Klein-Gordon field with mass $m_{\ell n}$.

\section{Numerical Results}

In this work we consider two simple couplings: i) $F_{1}(\phi ,\chi )=\chi
^{2}$ which results in a fourth-order potential $\chi ^{2}\Phi ^{2}$, and
ii) $F_{2}(\phi ,\chi )=\left( \phi \chi \right) ^{2}$ \ resulting in a
sixth-order potential $\left( \phi \chi \right) ^{2}\Phi ^{2}$. From Eq. (%
\ref{pot_sch}), the corresponding Schr\"{o}dinger potentials are%
\begin{eqnarray}
\hspace{-0.75cm}V_{1} &=&\frac{\ell ^{2}}{r^{2}}+4\mathcal{\eta }\left(
\frac{1}{s}-2\right) \frac{\left( \frac{r}{r_{0}}\right) ^{4\lambda s}}{%
\left[ \left( \frac{r}{r_{0}}\right) ^{4\lambda s}+1\right] ^{2}} \\
\hspace{-0.75cm}V_{2} &=&\frac{\ell ^{2}}{r^{2}}+4\mathcal{\eta }\left(
\frac{1}{s}-2\right) \frac{\left( \frac{r}{r_{0}}\right) ^{4\lambda s}\left[
\left( \frac{r}{r_{0}}\right) ^{4\lambda s}-1\right] ^{2}}{\left[ \left(
\frac{r}{r_{0}}\right) ^{4\lambda s}+1\right] ^{4}}
\end{eqnarray}

In order to investigate numerically the massive states, firstly we consider
the region near the origin ($r\ll r_{0}$). For $\lambda s\geq 1/2$, the
coupling functions $F_{1}$ and $F_{2}$ go to zero as $r^{4\lambda s}$ then
potentials are dominated by the contributions of the angular momentum
proportional to $1/r^{2}$,
\begin{equation}
V(r)\approx \frac{\ell ^{2}}{r^{2}},
\end{equation}%
and the nonsingular solutions\ in $r=0$ are
\begin{equation}
\rho _{n\ell }^{\left( 0\right) }(r)=J_{\ell }(m_{n}r).~,\ \ \ell \geq 0
\label{n4a}
\end{equation}
Hence, for each value of $\ell $, Eq. (\ref{n4a}) is used as an input for
the Runge-Kutta-Fehlberg method that produces a fifth order accurate
solution.

We now define the probability for finding scalar modes with mass $m_{\ell n}$
and angular moment $\ell $ inside the tube of radius $r_{0}$ as
\begin{equation}
P_{\ell n}=\frac{\int\limits_{r_{\text{min}}}^{r_{0}}dr~r\left[ \rho _{\ell
n}\left( r\right) \right] ^{2}}{\int\limits_{r_{\text{min}}}^{r_{\text{max}%
}}dr~r\left[ \rho _{\ell n}\left( r\right) \right] ^{2}}.
\end{equation}
Here $r_{min}\ll r_{0}$ is used as the initial condition and $r_{max}$ is
the characteristic box length used for the normalization procedure, being a
value where the Schr\"{o}dinger potentials are close to zero and where the
massive modes oscillate as plane waves.

\begin{figure}[]
\hspace{-0.2cm}\includegraphics[{angle=0,width=4.3cm}]{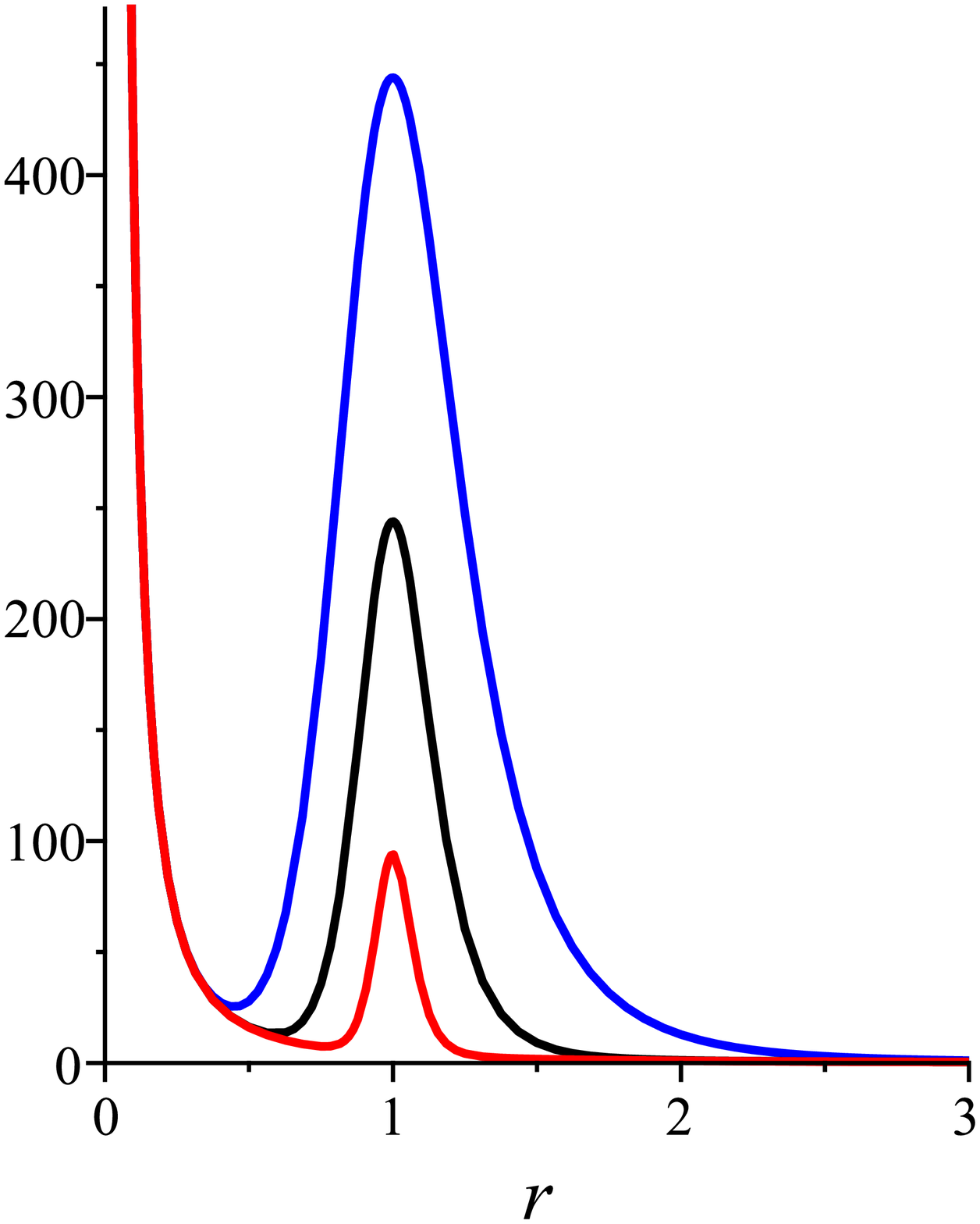}
\hspace{-0.2cm} \includegraphics[{angle=0,width=4.3cm}]{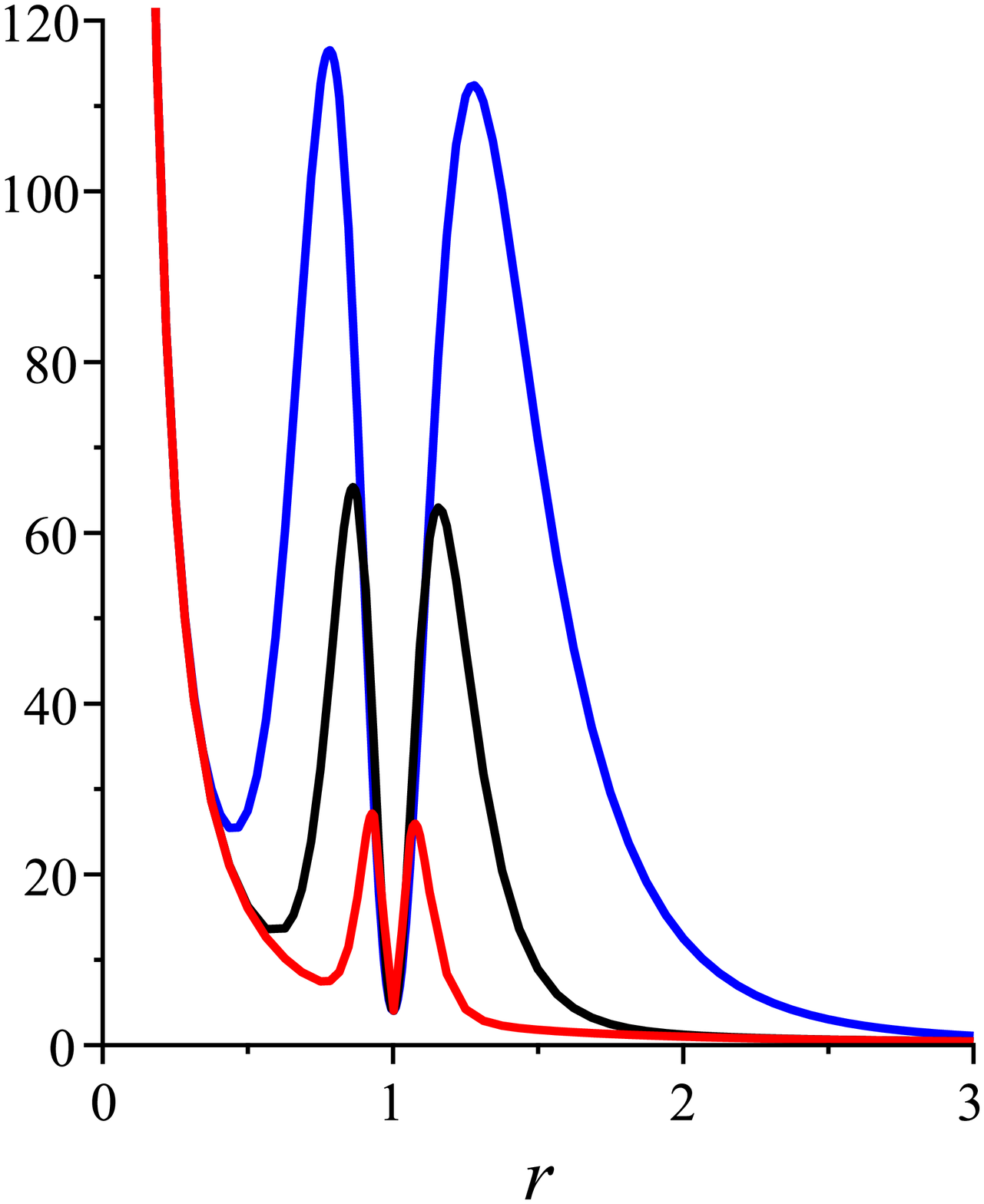}
\hspace{0.2cm}\includegraphics[{angle=0,width=4.3cm}]{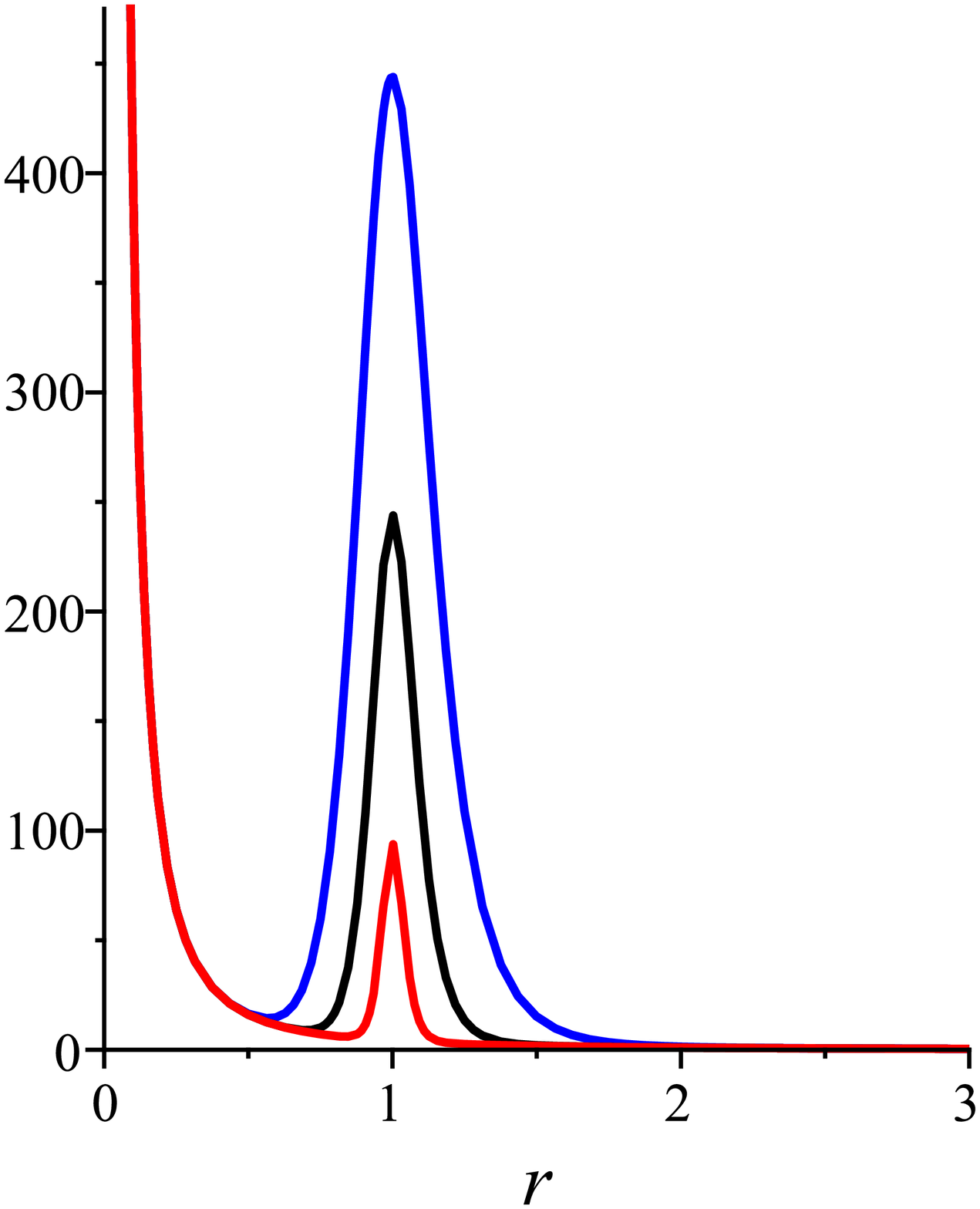}
\hspace{-0.2cm} \includegraphics[{angle=0,width=4.3cm}]{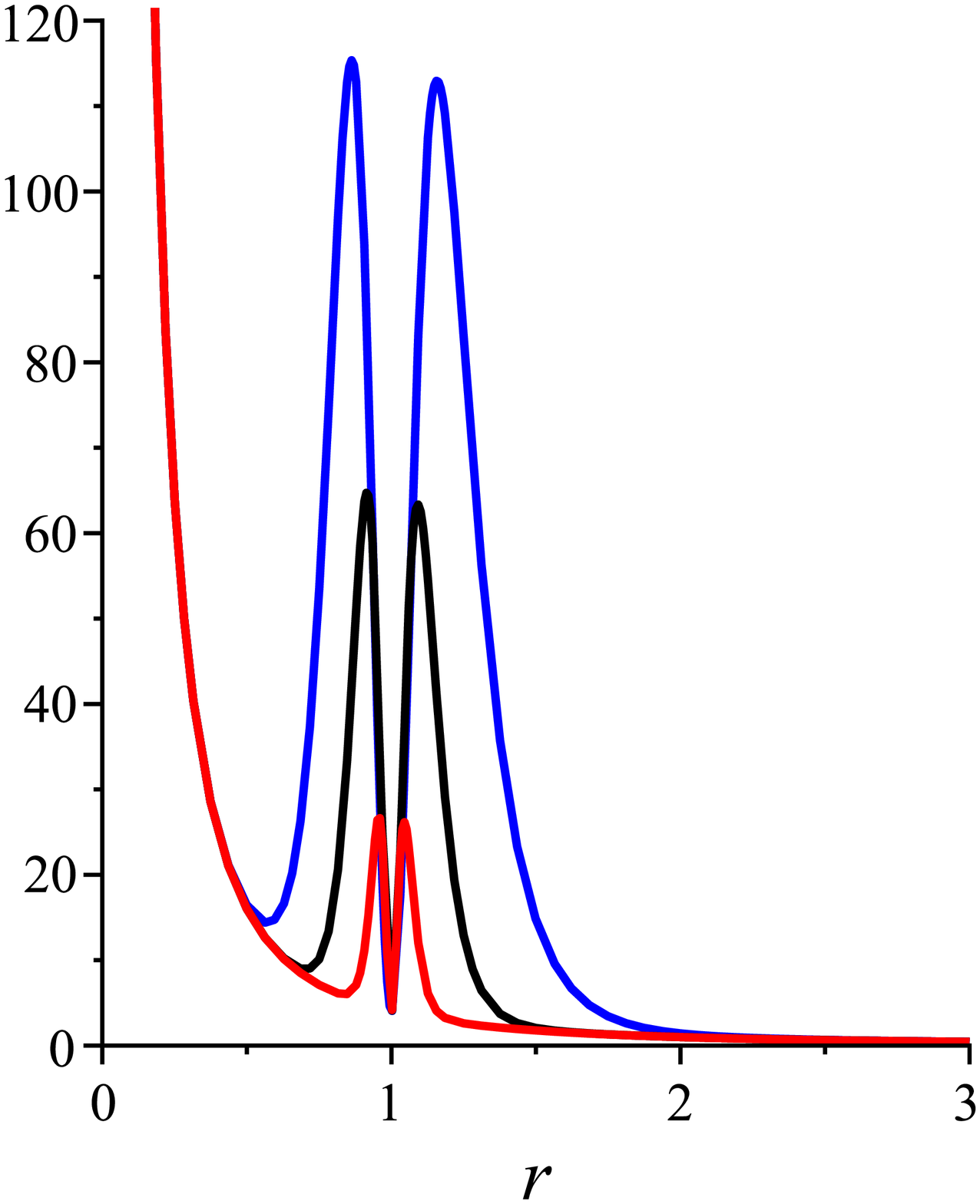}
\vspace{-0.15cm}
\caption{Schr\"{o}dinger-like potentials $V_{1}$ (left) and $V_{2}$ (right)
for $\ell =2$. We fix $r_0=1$, and $\protect\eta=30$. We have a) $\protect%
\lambda=30$ (upper figures) and b) $\protect\lambda=50$ (lower figures). In
all figures $s=0.06$ (blue), $s=0.1$ (black), $s=0.2$ (red). }
\label{potVsch_l}
\end{figure}

From the energy density considerations of Sec. \ref{tube}, larger values of $%
\lambda $ favor the existence of a Schr\"{o}dinger potential with structure
similar a tube\ barrier in $r=r_{0}$. \ The Fig. \ref{potVsch_l} depicts the
Schr\"{o}dinger-like potentials $V_{1}(r)$ and $V_{2}(r)$ for $\ell =2$, $%
\lambda =30$, $50$, and fixed $\eta =30$ and $r_{0}=1$. The potentials in
general diverge in $r\rightarrow 0$, assume a form of a barrier around $%
r=r_{0}$ (with a local maximum at $r=r_{0}$ for $V_1$ and a local minimum
for $V_2$) and asymptote to zero as $r\rightarrow \infty $, indicating the
possible presence of resonances. The increasing of $\eta $ turns the barrier
of the potential higher, whereas the increasing of $\lambda $ turns it
thinner. We noted that $\ell $ influences on the behavior of the potential
for $r<r_{0}$ but has no sensible influence on the barrier. We also observed
that the increasing of $r_{0}$ turns wider the potential barrier.


\begin{figure}[]
\includegraphics[{angle=0,width=4.2cm,height=5.0cm}]{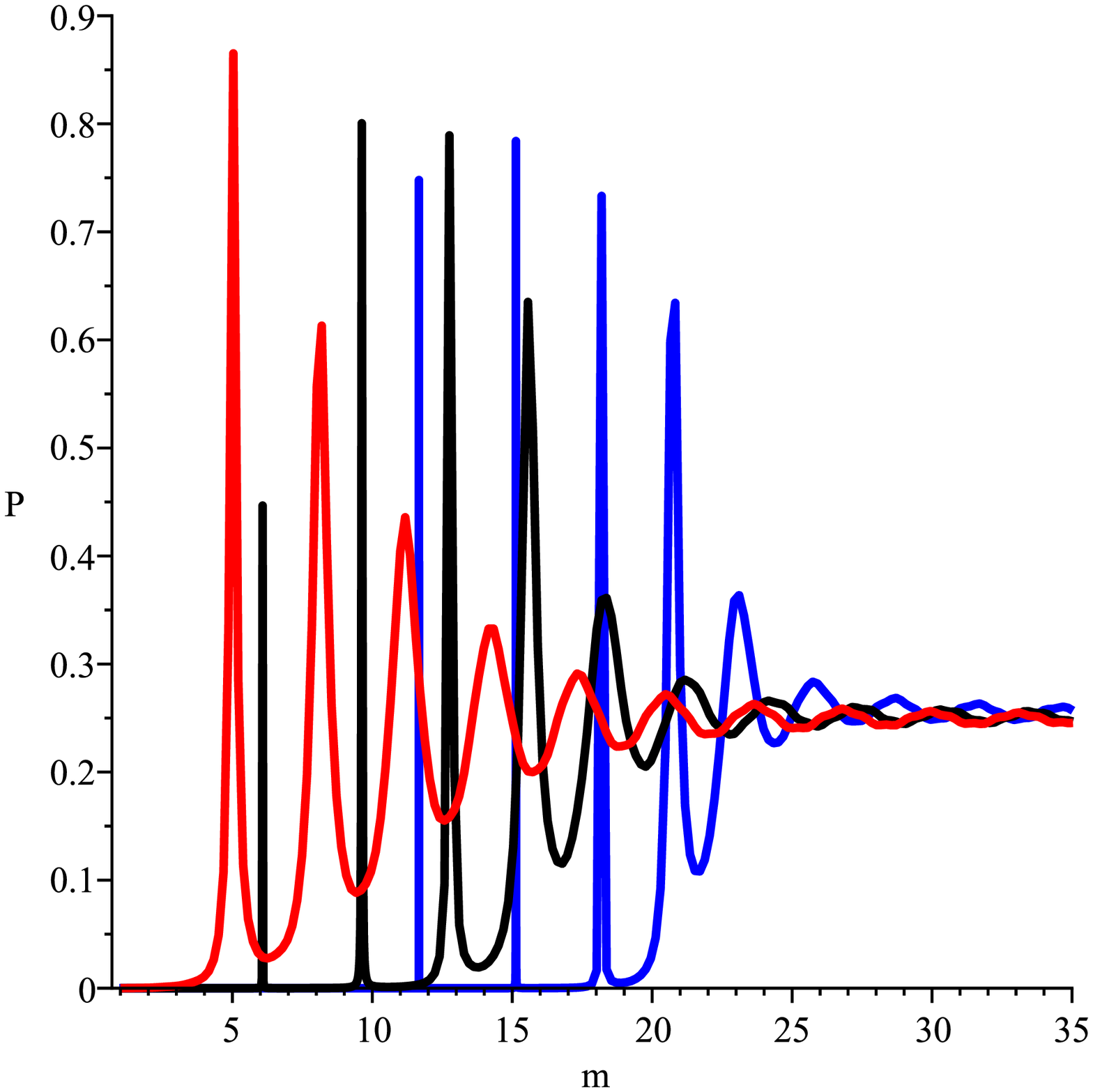}
\includegraphics[{angle=0,width=4.2cm,height=5.0cm}]{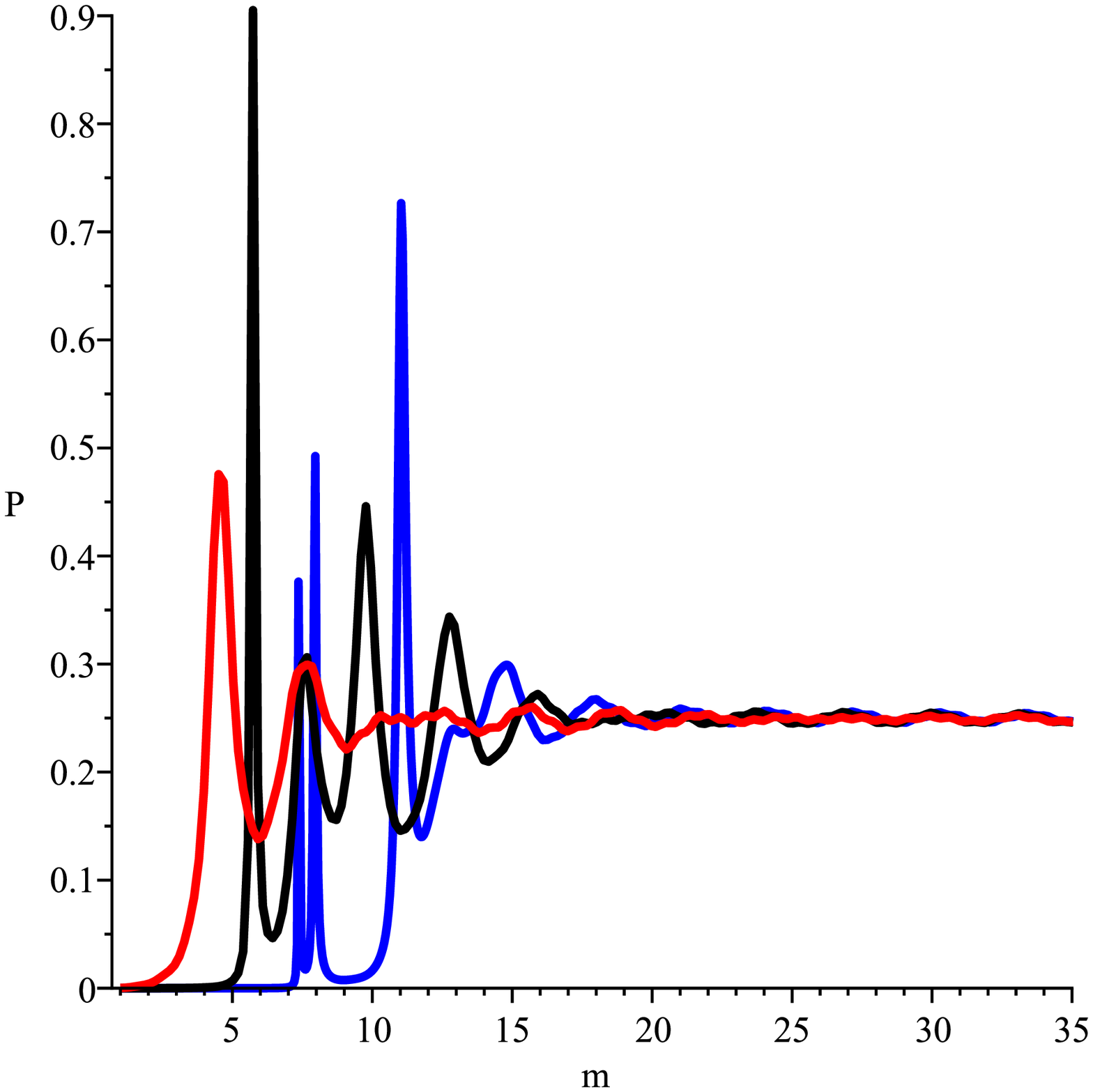}
\includegraphics[{angle=0,width=4.2cm,height=5.0cm}]{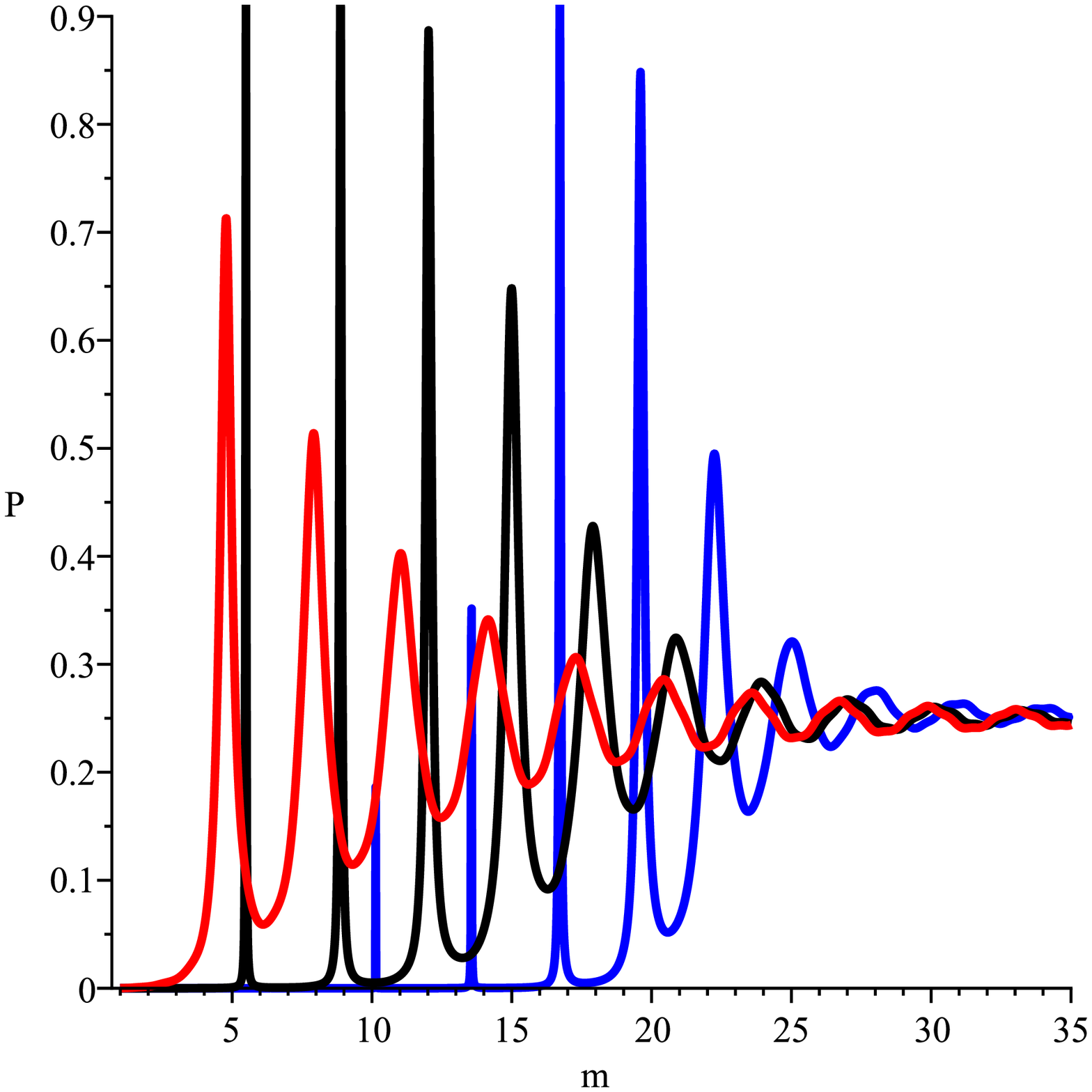}
\includegraphics[{angle=0,width=4.2cm,height=5.0cm}]{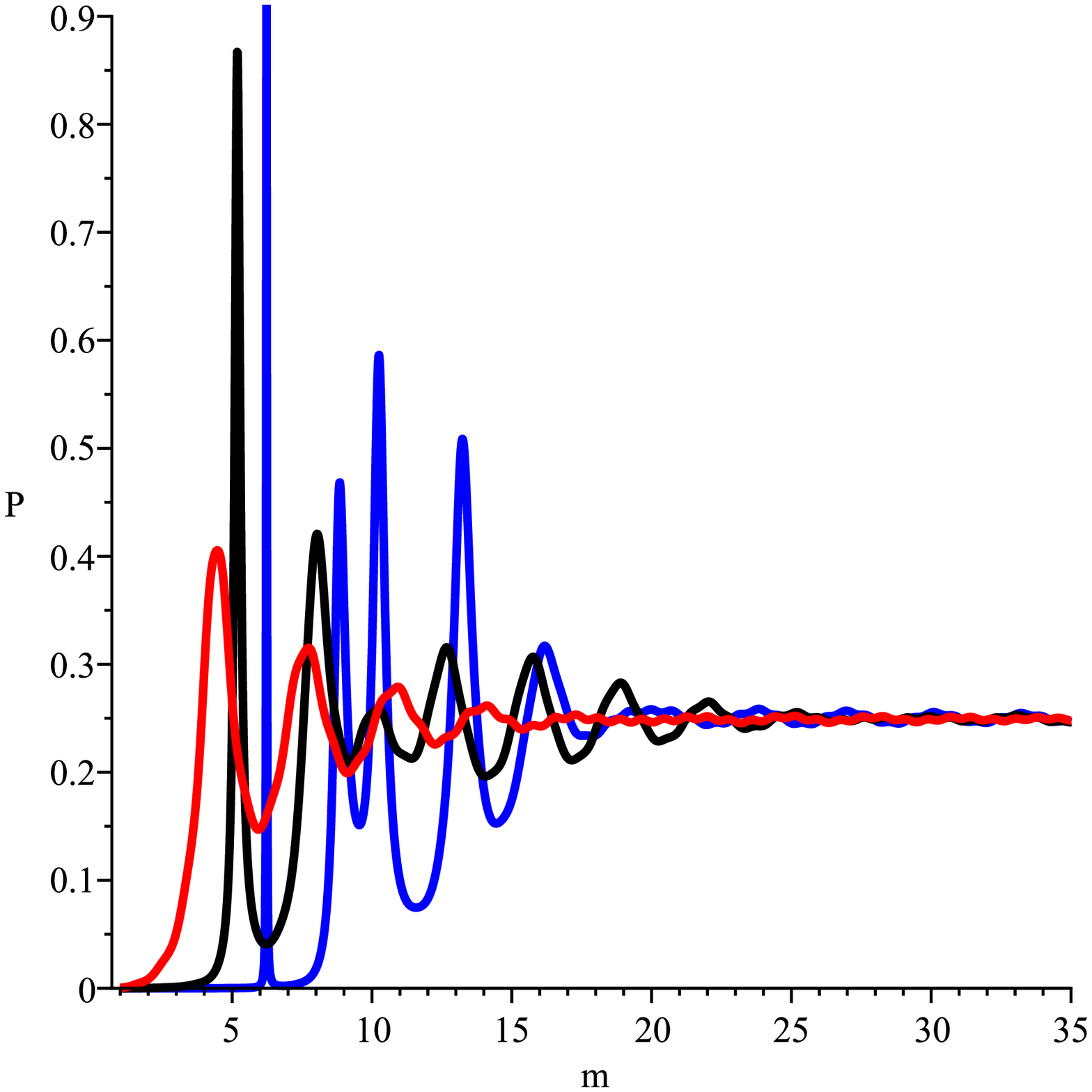}
\caption{$P_{\ell n}$ as a function of $m$ for couplings $F_1=\protect\chi^2$
(left) and $F_2=(\protect\phi\protect\chi)^2$ (right). We fix $\ell=2$, $%
\protect\eta=30$ and $r_0=1$. We have a) $\protect\lambda=30$ (upper
figures) and b) $\protect\lambda=50$ (lower figures). In all figures $s=0.06$
(blue), $s=0.1$ (black), $s=0.2$ (red).}
\label{Ptuberight}
\end{figure}

We remember that the field $\chi$ is responsible for the internal structure
of the defect, present for smaller values of the parameter $s$. This is
closely connected to the increasing of the height of the barrier in the
Schrodinger potentials, suggesting that lower values of $s$ are more
effective for trapping scalar particles. Figs. \ref{Ptuberight}a-b show some
results of $P_{\ell n}$ as a function of $m$ \ for couplings $F_1=\chi^2$
(left) and $F_2=(\phi\chi)^2$ (right). The plots are for $\ell =2$, $r_{0}=1$%
, $\eta =30 $ and for various values of $s$ and $\lambda$. We used $%
r_{min}=10^{-8}$, $r_{max}=4$ and step in $r$ equal to $\Delta r=0.002$. The
plots show several peaks of resonances, followed by a plateau for larger
masses where $P_{\ell n}=r_0/r_{max}$. The thinner is a peak, the larger is
the lifetime of the corresponding resonance. Comparing \ref{Ptuberight}a
(for $\lambda=30$) with \ref{Ptuberight}b (for $\lambda=50$), we note that,
for $\lambda\gg1$ larger values of $\lambda$ corresponds to slightly less
massive resonances, with lower lifetimes. This can be related to the thinner
barrier of the Schrodinger potential for larger values of $\lambda$. Also
from Figs. \ref{Ptuberight}a-b, we note that lower values of $s$ corresponds
to thinner peaks of resonance, agreeing with what expected from analysis of
the energy density and Schrodinger potential. The effect of the field $\phi$
in a direct coupling with $\Phi$ is to reduce the effectiveness of the
trapping mechanism. This is evident comparing left and right sides from
Figs. \ref{Ptuberight}a-b. We note that the resonances with coupling $%
F_1=\chi^2$ are better defined, in larger number and with larger masses in
comparison with those present when the coupling $F_2=(\phi\chi)^2$. We also
verified that the peak positions do not depend on the choice of $r_{max}$.

\section{Remarks and conclusions}

The present manuscript was firstly motivated by searching a first order
formalism in a braneworld scenario with two extra dimensions. We present it
in a  $(3,1)-$ dimensional system where the  $(y,z)-$ 
coordinates describe extra dimensions, in absence of gravitational
effects. The second purpose was to analyze the localization of scalar
particles living in a (1+1)-dimensional world (described by  $(t,x)-$ 
coordinates) interacting with a topological defect living in the extra
dimensions. In order to show explicitly such implementation, we have
considered an Abelian version of the color dielectric model described by two
coupled real scalar fields which have generated a tube-like topological
defect capable of trapping scalar particles.

Specifically, we have studied the localization of (1+1)-dimensional scalar
fields in a generalized tube-like topological defect whose cross-section is
a radial defect constructed with two scalar fields. The analysis  was performed 
by considering a general coupling between the defect and the scalar  field, 
and carefully  constructed to provide quantum mechanical description of the
amplitudes related to the scalar field modes. For the couplings $F(\phi
,\chi )=\left( \phi \chi \right) ^{2}$ $\ $and $F(\phi ,\chi )=\chi ^{2}$,
the numerical analysis of the Hamiltonian spectra showed that the field $%
\chi $, related to the presence of an internal structure on the defect, is
also the main responsible for the mechanism of trapping scalar particles
around the tube.  Further we must point out that a first-order
formalism for two extra dimensions was applied for a problem with axial
symmetry in $(3,1)$ dimensions, with interesting analytical simplifications
which turned the analysis much easier than a full numerical solution. 
The present analysis  can be considered as a startup in the 
construction of a first-order formalism for branesworlds  with two
extra dimensions. In this direction, the study of weak gravity field is 
currently being considered and results will be reported elsewhere. 

\section*{Acknowledgements}

The authors thank CAPES, CNPq and FAPEMA for financial support. RC and
AR Gomes thank MM Ferreira Jr. for discussions.


\begin{thebibliography}{99}
\bibitem{phenom} J. Greensite, Prog. Part. Nucl. Phys. 51 (2003) 1; T.
Suzuki, Nucl. Phys. Proc. Suppl. 30 (1993) 176; M. N. Chernodub,M. I.
Polikarpov, in  \textit{Confinement, Duality, and Nonperturbative Aspects
of QCD}, p. 387; M. N. Chernodub, V. I. Zakharov, Phys. Rev. Lett. 98
(2007) 082002; M. N. Chernodub, K. Ishiguro, A. Nakamura, T. Sekido, T.
Suzuki, V.I. Zakharov, PoSLAT2007, 174 (2007); M.N. Chernodub, Atsushi
Nakamura, V.I. Zakharov, Phys.Rev.D78, 074021 (2008).

\bibitem{cosm} A. Anabalon, S. Willison, J. Zanelli, Phys.Rev.D77, 044019
(2008); P. Mukherjee, J. Urrestilla, M. Kunz, A. R. Liddle, N. Bevis, M.
Hindmarsh, Phys.Rev.D83, 043003 (2011); M. Sakellariadou,
Lect.NotesPhys.738, 359 (2008); P.P. Avelino, C.J.A.P. Martins, C. Santos,
E.P.S. Shellard, Phys.Rev.Lett. 89, 271301 (2002); Erratum-ibid. 89, 289903
(2002); P.P. Avelino, L. Sousa, Phys.Rev.D83, 043530 (2011).

\bibitem{condmat} J.C.Y. Teo and C.L. Kane, Phys. Rev. B 82, 115120 (2010);
M.A. Silaev and G.E. Volovik, J. Low Temp. Phys, 161, 460 (2010); T. Fukui
and T. Fujiwara, \textit{Z2 index theorem for Majorana zero modes in a class
D topological superconductor}, arXiv:1009.2582; T.Sh. Misirpashaev and G.E.
Volovik, Physica, B 210, 338 (1995); G.E. Volovik, Pis'ma ZhETF 93, 69
(2011).

\bibitem{brane-origins} V.A. Rubakov and M.E. Shaposhnikov, Phys. Lett. B
125, 136 (1983); V.A. Rubakov and M.E. Shaposhnikov, Phys. Lett. B 125, 139
(1983); E.J. Squires, Phys. Lett. B 167, 286 (1986); M. Visser, Phys. Lett.
B 159, 22 (1985); K. Akama, Lect. Notes Phys. 176, 267 (1982); I.
Antoniadis, Phys. Lett. B 246, 377 (1990).

\bibitem{br1} K. Akama, Lect. Notes Phys. 176, 267 (1982).

\bibitem{br2} V. A. Rubakov and M. E. Shaposhnikov, Phys. Lett. B 125, 136
(1983).

\bibitem{br3} V. A. Rubakov and M. E. Shaposhnikov, Phys. Lett. B 125, 139
(1983).

\bibitem{br4} K. Akama, \emph{Gauge Theory and Gravitation}, Proceedings, edited by
K. Kikkawa, N. Nakanishi, and H. Nariai, Springer-Verlag, Nara, Japan (1983).

\bibitem{br5} M. Visser, Phys. Lett. B 159, 22 (1985).

\bibitem{thick1} O. DeWolfe, D. Z. Freedman, S. S. Gubser and A. Karch,
Phys. Rev. D 62, 046008 (2000).

\bibitem{thick2} M. Gremm, Phys. Lett. B 478, 434 (2000).

\bibitem{thick3} M. Gremm, Phys. Rev. D 62, 044017 (2000).

\bibitem{thick4} A. Kehagias and K. Tamvakis, Mod. Phys. Lett. A 17, 1767
(2002).

\bibitem{thick5} C. Csaki, J. Erlich, T. Hollowood and Y. Shirman, Nucl.
Phys. B 581, 309 (2000).

\bibitem{thick6} A. Campos, Phys. Rev. Lett. 88, 141602 (2002).

\bibitem{thick7} R. Guerrero, A. Melfo and N. Pantoja, Phys. Rev. D 65,
125010 (2002).

\bibitem{thick8} D. Bazeia, C. Furtado and A. R. Gomes, J. Cosmol.
Astropart. Phys. 0402 (2004) 002.

\bibitem{thick9} D. Bazeia and A. R. Gomes, J. High Energy Phys. 05 (2004)
012.

\bibitem{thick10} D. Bazeia, F. A. Brito and A. R. Gomes, J. High Energy
Phys. 0411 (2004) 070.

\bibitem{thick_rev} V. Dzhunushaliev, V. Folomeev and M. Mina- mitsuji, Rep.
Prog. Phys. 73, 066901 (2010).

\bibitem{field-localiz} Y. X. Liu, J. Yang, Z. H. Zhao, Chun-E Fu and Y. S.
Duan, Phys. Rev. D 80, 065019 (2009); Y. X. Liu, H. T. Li, Z. H. Zhao, J. X.
Li and J. R. Ren, JHEP 0910, 091 (2009); I. Oda, Phys. Lett. B 496, 113
(2000); C.A.S. Almeida, R. Casana, M.M. Ferreira and A.R. Gomes, Phys. Rev.
D 79, 125022 (2009).

\bibitem{bmm1} D. Bazeia, J. Menezes, and R. Menezes, Phys. Rev. Lett. 91,
241601 (2003).

\bibitem{bmm2} D. Bazeia, J. Menezes, R. Menezes, Mod. Phys. Lett. {B19},
801 (2005).

\bibitem{vortex} H. B. Nielsen and P. Olesen, Nucl. Phys. B 61, 45 (1973).

\bibitem{nonabelian} M. Shifman and A. Yung, Rev. Mod. Phys. 79, 1139
(2007). M. Eto, T. Fujimori, S. B. Gudnason, Y. Jiang, K. Konishi, M. Nitta
and K. Ohashi, JHEP 1011 (2010) 042. R. Auzzi, S. Bolognesi, J. Evslin, K.
Konishi and A. Yung, Nucl. Phys. B673, 187-216 (2003). M. Eto, T. Fujimori,
Y. Isozumi, M. Nitta, K. Ohashi, K. Ohta and N. Sakai, Phys. Rev. D 73,
085008 (2006). J. Evslin, Kenichi Konishi, Muneto Nitta, Keisuke Ohashi, W.
Vinci,  \emph{Non-Abelian Vortices with an Aharonov-Bohm
Effect}, ArXiv 1310.1224 [hep-th].

\bibitem{er} R. Emparan and H.S. Reall, Class. Quant. Grav., 23, R169, 2006;
ibid, Living Rev.Rel. 11:6,2008.

\bibitem{rt} V.A. Rubakov and A.N. Tavkhelidze, Phys. Lett. B 165, 109
(1985).

\bibitem{rub} V. A. Rubakov, Prog. Theor. Phys. 75, 366 (1986).

\bibitem{kks} B. Kleihaus, J. Kunz, and Y. Shnir. Phys. Rev. D 68, 101701
(2003); Phys. Rev., D 70, 065010 (2004).

\bibitem{tp} Gerard \'t Hooft, Nucl. Phys. B 79, 276 (1974); A. M. Polyakov.
JETP Lett. 20, 194 (1974).

\bibitem{kkl} B. Kleihaus, J. Kunz, M. Leissner, Phys. Lett. B 663, 438
(2008).

\bibitem{rub2} V. A. Rubakov, \textit{Classical Theory of Gauge Fields},
Princeton Univ. Press, Princeton, 2002.

\bibitem{km} F. R. Klinkhamer and N. S. Manton, Phys. Rev. D 30, 2212 (1984).

\bibitem{rv} E. Radu and M. S. Volkov, Phys. Rept. 468, 101 (2008).

\bibitem{fried} R. Friedberg and T. D. Lee, Phys. Rev. D15, 1694 (1977); 16,
1096 (1977); 18, 2623 (1978).

\bibitem{wil} L. Wilets, Nontopological Solitons (World Scientific,
Singapore, 1989).

\bibitem{scalar-localiz} B. Bajc and G. Gabadadze, Phys. Lett. B 474 (2000)
282. Heng Guo, A. Herrera-Aguilar, Yu-Xiao Liu, D. Malagon-Morejon, R. R.
Mora-Luna, Phys.Rev. D87 (2013) 095011. Heng Guo, Yu-Xiao Liu, Zhen-Hua
Zhao, Feng-Wei Chen, Phys.Rev. D85 (2012) 124033.

\bibitem{raj1} R. Hobart, Proc. Phys. Soc. Lond.82, 201(1963).

\bibitem{raj2} G. H. Derrick J. Math. Phys. 5, 1252 (1964).

\bibitem{raj3} R. Rajaraman, \textit{Solitons and Instantons},
North-Holland, Amsterdan, 1982.

\bibitem{color1a} M. Alford, J.A. Bowers, and K. Rajagopal, Phys. Rev. D 63,
074016 (2001).

\bibitem{color1b} M. G. Alford, K. Rajagopal, T. Schaefer, A. Schmitt,
Rev.Mod.Phys.80, 1455 (2008).

\bibitem{color2a} R.Casalbuoni, R. Gatto, M. Mannarelli and G. Nardulli,
Phys. Lett. B 511, 218 (2001);

\bibitem{color2b} R.Casalbuoni, R. Gatto, M. Mannarelli and G. Nardulli,
Phys.Rev. D66 (2002) 014006.

\bibitem{deconf} F. Sannino, N. Marchal, W. Sch\"afer, Phys.Rev. D66 (2002)
016007.

\bibitem{br-inta} N. Arkani-Hamed, S. Dimopoulos, G. Dvali, and N. Kaloper,
Phys. Rev. Lett. 84, 586 (2000);

\bibitem{br-intb} N. Kaloper, Phys. Lett. B 474, 269 (2000).

\bibitem{n-coma} P.-M. Ho and Y.-T. Yeh, Phys. Rev. Lett. 85, 5523 (2000).

\bibitem{n-comb} O. Bertolami and L. Guisado, J. High Energy Phys. 12, 013
(2003).

\bibitem{dobr} R.L. Dobrushin, Theor. Prob. Appl.17, 582 (1972).

\bibitem{bor} C. Borgs, Commun. Math. Phys.96, 251 (1984)

\bibitem{bsr} D. Bazeia, M.J. dos Santos and R.F. Ribeiro, Phys. Lett. A208,
84 (1995).

\bibitem{igg} A. Alonso Izquierdo, M.A. Gonzalez Leon, J. Mateos Guilarte,
Phys.Rev. D65 (2002) 085012.
\end{thebibliography}
\end{document}